# Defect-related photoluminescence of hexagonal boron nitride

**Luc MUSEUR**

*Laboratoire de Physique des Lasers – LPL, CNRS UMR 7538, Institut Galilée, Université Paris 13, 93430 Villetaneuse, France*

**Eduard FELDBACH**

*Institute of Physics, University of Tartu, 142 Riia Str., 51014 Tartu, Estonia*

**Andrei KANAEV**

*Laboratoire d'Ingénierie des Matériaux et des Hautes Pressions – LIMHP, CNRS, Institut Galilée, Université Paris 13, 93430 Villetaneuse, France*

# Abstract.

Photoluminescence of polycrystalline hexagonal boron nitride (hBN) was measured by means of time- and energy-resolved spectroscopy methods. The observed bands are related to DAP transitions, impurities and structural defects. The excitation of samples by high-energy photons above 5.4 eV enables a phenomenon of photostimulated luminescence (PSL), which is due to distantly trapped CB electrons and VB holes. These trapped charges are metastable and their reexcitation with low-energy photons results in anti-Stockes photoluminescence. The comparison of photoluminescence excitation spectra and PSL excitation spectra allows band analysis that supports the hypothesis of Frenkel-like exciton in hBN with a large binding energy.

# I. Introduction.

Optical and fluorescence properties of hexagonal boron nitride deserved particular interest during last decade since the first observation of an intense far-UV exciton emission[1,2] making this material a candidate for use in new light-emitting devices. A successful synthesis of high-purity single crystal samples initially at high-P,T [1,2] and later at atmospheric conditions [2] enabled this achievement. Accordingly, understanding of the electronic band and exciton structure of hBN becomes even more important issue as it serves a basic system for single-wall [3,4] and multi-wall boron nitride nanotubes[5-9].

Despites of many efforts in the past, the electronic properties of hBN remain largely unknown. They have been earlier studied by luminescence [1,10-17], optical reflectance and absorption [18-21], photoconductivity[22], x-ray emission [23-25], inelastic x-ray scattering [26-28], x-ray absorption [25,29,30], or electron energy loss [31-33] spectroscopy. After all, a large spread of band gap energies reported in literature, ranging from 3.1 to 7.1 eV[14], is currently explained by sample quality and related to experimental methods used. Recently, arguing on generally high luminosity of the free exciton luminescence, Watanabe *et al* [1] assumed that hBN is a direct band gap material. They have measured the band gap energy of 5.971 eV and inferred an exciton binding energy of 0.149 eV that corresponds to the Wannier exciton model. However this result is in a large disagreement with the most recent theoretical calculations using the so-called all electrons GW approximation [34,35]. They predict hBN to be an indirect band gap material with a band gap energy of 5.95 eV and a lowest direct interband transitions at 6.47 eV [34]. Moreover, Arnaud *et al.* [34] has deduced a huge exciton binding energy of 0.72 eV and predicted that the low-lying exciton in hBN belongs to the Frenkel type. The intense free exciton luminescence observed in single crystal hBN is explained by a large oscillator strength of merged excitonic transitions [34].

In view of many disagreements further confrontation between experiment and theory will continue. In these conditions, more experimental data concerning electronic and related optical properties of hBN are highly required. In particularly, energy- and time-resolved photoluminescence methods provide valuable information about excitonic and interband transitions. Moreover, in polycrystalline samples energy transfer to impurities and defects may inhibit these intrinsic transitions and strongly affect the fluorescence spectra. However, the energy transfer is specific to excitation energy and the relevant excited states can be experimentally resolved.

In the present article we report on detailed analysis of the defect-related intraband luminescence of hBN. Important information about intrinsic properties of hBN is obtained by combining two experimental approaches: time- and energy- resolved photoluminescence and photostimulated luminescence. The discussion is organized as follow. The experiment is described in section II and experimental results are presented on the section III. In section IV-A we discuss the nature of the observed luminescence bands. Finally, in section IV-B, careful comparison between photoluminescence excitation (PLE) and photostimulated luminescence (PSL) excitation spectra brings more precision to exciton and bandgap transitions of hexagonal BN.

## II. Experiment.

The samples were prepared from commercial hexagonal BN powders (Alfa 99.5%) compacted in pellets of size 8x8x1 mm$^3$ at hydrostatic pressure of 0.6 GPa. The grit size of the hBN powder has been estimated by means of granulometry and transmission electron microscopy (JEM 100C JEOL). It ranged from 0.3 to 10 $\mu$m with an average particle size of 3.1 $\mu$m corresponding to the maximum in the mass distribution curve. The samples were then heated at 800 K under vacuum for 12 hours to avoid organic impurities and traces of water.

The luminescence properties of the samples were studied using VUV synchrotron-radiation (SR) source of DORIS storage ring at HASYLAB (DESY, Hamburg). The facility of the SUPERLUMI station used in the experiments is described in details elsewhere [36]. Briefly, samples were cooled down to 8K and irradiated by monochromatized SR ($\Delta\lambda = 3.3 \overset{o}{A}$) under high vacuum (~$10^{-9}$ mbar). The measurements of luminescence spectra were carried out using a visible 0.275-m triple-grating ARC monochromator equipped with a CCD detector or a photomultiplier operating in the photon-counting mode. The pulse structure of SR (130 ps, 5 MHz repetition rate) enables time-resolved luminescence analysis at time-scale of 200 ns with sub-nanosecond temporal resolution. Spectra were recorded within a time gate $\Delta\tau$ delayed after the SR excitation pulse. Typically two time gates have been used simultaneously: a fast one of $\Delta\tau_1 = 1-4\,ns$ and a slow one of $\Delta\tau_3 = 22-200\,ns$. Complementary, luminescence decay curves were measured at fixed excitation and luminescence photon energies. The recorded spectra were corrected for the primary monochromator reflectivity and SR current.

The photostimulated luminescence (PSL) excitation spectra of pd-hBN samples were measured at the BL 52 beamline of MAX-lab synchrotron (Lund Sweden). The experiment built up by the Tartu group is described in ref [37, 38]. At each excitation energy $E_{exc}$ the sample was irradiated by a fixed number of UV or VUV photons (2.5·$10^6$ counts) in the energy region of $E_{exc}$ = 5-15 eV. After completing the dose, the irradiation was stopped and the phosphorescence decay was measured. When the phosphorescence intensity drops until almost zero (PM noise level) that typically happens after few minutes, the sample was activated by a bright spectroscopic source at h$\nu$=2.0±0.5 eV through a double-prisms monochromator DMR-4 and the time dependence of PSL intensity was recorded. The integral PSL intensity was taken as a measure of the number of recombined electron-hole pairs.

# III. Results.

Under photoexcitation above 6 eV the luminescence spectra of hBN are composed of several luminescence bands (Figure 1). Basically, one broadband and one structured emissions between 3 and 4 eV, and two relatively narrow near bandgap emissions at 5.3 eV and 5.5 eV were distinguished. Two high-energy emissions were discussed in details in our recent publication[17]. They were assigned respectively to quasi donor-acceptor pairs (q-DAP) (5.3 eV band) and bound excitons (5.5 eV band). Below we complete the assignment of low-energy luminescence.

## A – Luminescence around 4 eV in hBN powder samples.

The hBN powder sample is a strongly luminescent material, which luminescence spectra depend on excitation energy. The low-temperature luminescence spectra of hBN samples recorded with different excitation energies varying from 4 eV to 6.5 eV are presented in Figure 1. Under excitation below 5.0 eV a strong structured UV emission is observed. Four peaks, labeled (α), (β), (γ) and (δ), are clearly visible. A multiple Gaussian fit procedure results in the peak energies of 4.099 eV (α), 3.912 eV (β), 3.731 eV (γ) and 3.539 eV (δ). These four peaks are equally spaced in energy by $\omega_g$ = 186 ± 1.4 meV. At room temperature these peaks are broadened, but no appreciable shift was detected. When the excitation energy exceeds 5.0 eV a very broad band (ΔE~1 eV) centered at 3.9 eV appears and superimposes with the structured emission (Figure 1.c and d). However, the peaks of the structured emission can always been observed on the top of the broad band.

At excitation energy of 4.27 eV only the structured UV emission can be observed. The luminescence decay curves of its four peaks (α)-(δ) are shown in Figure 2 by curve (ε). At this excitation energy the broad band dominates the luminescence spectra. Although the probed energy $E_{lum}$ = 3.91 eV is in coincidence with the peak (β) of the structured emission, the decay curve appears to be multi-exponential and longer lifetime than that of the peak (β).

The time-resolved photoluminescence method allows separating several contributions of superposed luminescence bands if their respective lifetimes are noticeably different. This is the case of the structured and continuous broadband emissions around 4 eV. Figure 3.a shows the total PLE spectrum measured at $E_{lum}$ = 3.73 eV. The last corresponds to the peak ($\gamma$) of the structured emission and is close to the maximum of the broadband emission. Two principal features can be distinguished in this spectrum: (i) a sharp onset at 4.09 eV followed by a decrease of the luminescence intensity until nearly zero at $E_{exc}$ = 4.9 eV, and (ii) a strong peak at 5.95 eV preceded by a weak structured footer that begins at 5.0 eV. The PLE spectra obtained with short and long time windows are displayed in Figure 3.b and c. The long time-window of $\Delta\tau_2 = 22-200\,ns$ selects long-lived excited states, which apparently are responsible for the broad emission band (Figure 3.c). The peak maximum at 5.95 eV strongly contributes to excitation of this band. This peak is blue-shifted with respect to that of the exciton absorption at 5.82 eV in hBN single crystal [1]. Nevertheless, a weak luminescence at 5.9 eV is also reported in hBN single crystal at low temperatures and assigned to another exciton series. The Stokes shift with respect to the excitation peak at 5.95 eV could results from inhomogeneities of the local field or dislocations such as stacking faults [1]. Therefore we tentatively assign the peak at 5.95 eV to the excitation of higher exciton series. This excitonic peak is preceded by a weak and unstructured footer ranging from 5 eV to 5.5 eV and is followed by a broad continuum until 7.6 eV. Conversely, the PLE spectra obtained with the short time window ($\Delta\tau_1 = 1-4\,ns$) show contributions from both short and long decay luminescence components (Figure 3.b). Nevertheless, a comparison with the PLE spectra of the long component (Figure 3.c) allows firmly assigning the structured emission with the onset at 4.09 eV to a short-lived excited state. Moreover, we note a second weak excitation onset of the structured emission at $E_{exc}$ = 5.2 eV.

Similar luminescence spectra to those depicted in Figure 1 have been recently reported on photoluminescence and cathodoluminescence experiments of commercial hBN powder [10, 39-48]. However, in these publications, no distinctions were made between the broadband and structured emissions. We now assign the structured UV emission to impurities (probably, C) and the broadband emission to deep donor-acceptor pair (DAP) of strongly localized center. These assignments will be discussed below in section IV-A.

**B – Photostimulated luminescence in hBN powder samples.**

The photostimulated luminescence arises from the trapping of free charge carriers in distant lattice sites with subsequent recombination of carriers released from these traps by a visible light [38]. The charge separation and trapping can result from different excitation processes listed below and depicted in Figure 4:

$$A^- + D^+ + h\nu_{exc} \rightarrow A^- + D^+ + e^- + h^+ \rightarrow A^o + D^o \quad (1)$$

$$A^- + D^+ + h\nu_{exc} \rightarrow A^o + D^+ + e^- \rightarrow A^o + D^o \quad (2)$$

$$A^- + D^+ + h\nu_{exc} \rightarrow A^- + D^o + h^+ \rightarrow A^o + D^o \quad (3)$$

$$A^- + D^+ + h\nu_{exc} \rightarrow A^o + D^o \quad (4)$$

The band-to-band transitions (Equation 1) or impurity-band ionization (Equations 2-3) are the most likely contributions to PSL. In these cases at least one of photoexcited charge carriers (e⁻ or h⁺) becomes free and can migrate away from the point of excitation before being trapped by the acceptor or donor defects. Spatially closely trapped charges then annihilate giving rise to phosphorescence. In contrast, at rather large distance their lifetime becomes infinite and their recombination can only be possible after reactivation by light or thermally. The Figure 5 shows the phosphorescence decay curves ($E_{lum} = 3.91\,\text{eV}$) for different photon excitation energies ranging from 5.4 eV to 5.9 eV. At $E_{exc} \leq 5.50$ eV the phosphorescence decay curves are characterized by an almost mono-exponential decay with characteristic time of $\tau = 5.8$ s (the dotted line in Figure 5 shows the PM dark noise). However, at $E_{exc} \geq 5.60$ eV another extremely long-lived component of low intensity appears which looks like a plateau in Figure 5. We ascribe this long component to the recombination of charges trapped at large distance tending to infinity.

The observed phosphorescence might also be related to the "dark" (dipole forbidden) exciton states theoretically predicted in hBN by Wirtz et al. [35]. The authors suggested a coupling between the "white" (dipole allowed) and "dark" exciton states, which makes them appear in the absorption spectra of hBN monocrystals reported by Watanabe *et al* [1]. However, we disregard their contribution in PSL experiments since the phosphorescence is already observed at $E_{exc} = 5.1$ eV (Figure 5), which is well below the exciton energy in hBN. Moreover, PSL is related to the states with infinite lifetime (excited above 5.5 eV), which depopulation is only triggered by light. In contrast, the hypothetic "dark" states become accessible when the crystalline symmetry is broken through defects or limited sample quality: this is our case, since no free exciton emission has been observed. Accordingly, the singlet and triplet exciton are expected to be strongly coupled in our polycrystalline hBN samples and their lifetime is short (sub-nanosecond). In this case no long-lived exciton states are expected.

The PSL excitation spectrum is represented by open circles in Figure 6. The full measured spectrum is shown in the insert of the figure for indicative purposes. In the following, we will discuss the PSL phenomenon specific to excitation energies ranging from 5 to 7 eV. The signal shows up appreciably at $E_{exc} > 5.50$ eV in correlation with the appearance of a long component of the phosphorescence decay (Figure 5). The PSL grows almost exponentially with the excitation energy until $E_{exc} \equiv E_{i1} = 5.7\,eV$ where it locally attains a maximum. At this energy the slope suddenly changes and the growth of the signal becomes much slower. Another local maximum of the PSL spectrum appears at $E_{exc} \equiv E_{i2} = 6.10$ eV. Finally, for excitation energies above 6.2 eV the PSL signal increases progressively until 7 eV where a plateau is reached.

For the sake of the discussion we have superimposed in Figure 6 the PLE spectrum of the near band edge luminescence at $E_{lum} = 5.3\,\text{eV}$. This emission was discussed in details in our previous publication[17] and has been assigned to the radiative recombination of the so called quasi donor acceptor pair (q-DAP).

# IV. Discussion.

### A – Luminescence of polycrystalline hBN samples

The structured UV luminescence (Figure 1.a and b) has been reported in the past by numerous groups, either in cathodoluminescence [10, 43, 46, 47] or photoluminescence experiments [10, 14, 39-42, 44, 45, 47, 48]. Some authors have assigned it to intrinsic luminescence of hBN molecular layers [47] or to phonons assisted band edge luminescence [42]. However, recent experiments [1] and theoretical calculations [34] indicate the hBN bandgap energy in the range from 6 to 6.5 eV that disables these interpretations. Others authors [45, 49, 50] claimed that the structured luminescence is due to transition between the conduction band and acceptor carbon atom at substitutional N site. Recently, experimental evidence of the correlation between carbon and oxygen contents in hBN samples and intensity of the structured luminescence has been given[51].

In a recent publication [39] the structured luminescence of hBN has been attributed to defects or impurities. The peaks (β), (γ) and (δ) in Figure1.a were ascribed to phonon replicas of the zero phonon line (α) involving TO phonons ($\omega_{TO}$ = 169 meV). Due to the low sample temperature in present experiments, the luminescence peaks become well resolved, allowing more precise determination of their spectral positions than in room-temperature experiments [39, 42]. The obtained phonon energy of $\omega_g$ = 186 meV, which falls between the known energy of LO (199 meV) and degenerated TO/LO (169 meV) phonons at the Γ point [18, 19, 26, 52], corresponds more probably to a local phonon mode around the impurity involved in the luminescence process. Moreover, the pronounced red shoulder observed on each phonon replica peak shows that multi-phonon processes play a major role in the energy relaxation of the impurity. The coupling between the defect and the lattice is weak as it is shown by the Huang-Rhys factor S = 1.3 obtained from the normalized peak intensities. This assignment is consistent with fast luminescence decay of the peaks (α), (β), (γ) and (δ) (figure 2).

The photoluminescence excitation (PLE) spectrum shown in Figure 3.b supports our interpretation of the defect related luminescence. Indeed, it reflects features characteristic of such transitions. The luminescence intensity steeply increases and then decreases progressively with the excitation energy. The sharp onset in the PLE spectrum at 4.09 eV corresponds to the minimum energy required to excite the impurity. As in the case of weak electron-phonon coupling, this absorption onset coincides with the zero-phonon luminescence peak (α). Interestingly, two phonons replicas separated by $\omega_e$ =170 meV, that is close to the energy of the TO/LO degenerated phonons[26], can be observed in the PLE spectra. They are indicated by arrows in Figure 3.a. We see that the energies of phonons involved in the luminescence ($\omega_g$) and excitation ($\omega_e$) processes are different. This results from the electronic state of the defect, which is not the same in both cases. Consequently, lattice deformations around the defect are different that affects the local phonon frequencies. The fact that $\omega_g > \omega_e$ signifies a stronger matrix interaction with the impurity in the excited state.

We do not evidence the impurity involved in the luminescence process in current experiments. However, we can guess about the influence of carbon with more or less confidence. Recent experiments reveal that the structured luminescence strongly shows up in hBN samples contaminated by carbon and oxygen[51]. This result is consistent with our previous results[39] and is supported by our complementary photoluminescence experiments carried out with a pyrolytic BN sample (pBN). These experiments will be described in details in a forthcoming article[53]. The pBN is a high purity material, free of carbon compound, obtained by gas-phase reactions between $BCl_3$ and $NH_3$ at 2300 K and deposition on a Si substrate (CVD method). Actually, pBN sample exhibits no structured emission around 4 eV whatever the excitation energy is. Its excitation by energy photon above 4.5 eV uniquely results in a continuous emission band similar to that observed in hBN[53]. Moreover, the PLE spectrum of hBN (Figure 3) is rather complicated with a first edge at 4.09 eV and a second one at 5.2 eV. This may indicate that the defect involved in the radiative process has at least two excited levels separated by 1.1 eV positioned within the bandgap. Since, the substitutional carbon impurity at nitrogen site ($C_N$) is supposed to introduce two energy levels split by ~0.8 eV in the energy range 3.2 – 4.9 eV below the conduction band of hBN, [50] its participation in the luminescence process is most likely.

We discuss now the nature of the broad luminescence band observed in hBN powder sample under photoexcitation above 5 eV. Due to its long and multi exponential decay (curve (ε) in figure 2), we assign this luminescence to radiative recombination of deep donor and acceptor pairs (DAP):

$$A^o + D^o \rightarrow A^- + D^+ + h\nu_{DAP} \qquad (5)$$

We remark that this deep DAP is different from that related to the 5.3-eV emission (q-DAP[17]). As Figure 3.c shows, the correspondent PLE spectrum of the long-lived luminescence component is dominated by the excitonic peak at 5.95 eV. This fact indicates that the energy transfer to the DAP recombination channel via free excitons is efficient:

$$\begin{aligned} hBN + h\nu_{exc} &\rightarrow exciton \\ exciton + A^- + D^+ &\rightarrow A^0 + D^0 \end{aligned} \qquad (6)$$

This excitonic peak is preceded by a weak shoulder for excitation energies ranging from 5 eV to 5.5 eV. We have assigned this shoulder to a direct excitation of the deep DAP (Equation 4). The PSL spectrum displayed in Figure 6 shows that no noticeable photostimulated luminescence can be observed following hBN irradiation between 5 and 5.5 eV that indicates no efficient population of distant traps. In contrast to the direct ionization of the donor or acceptor (Equations 2-3), the mechanism (4) does not lead to the charge injection into CB ($e^-$) or VB ($h^+$) and does not contribute significantly to PSL.

A very large bandwidth of the deep DAP band (~1 eV) and its nearly symmetrical shape, suggest that at least one of defects involved in the emission is strongly coupled to the lattice[54]. This assumption is supported by a comparison of the luminescence spectra obtained at room and low (9 K) temperatures: when the temperature is increased from 9 K to 300 K (Figure1.c) thermally activated quenching is intensified. The intensity of the broad DAP band then drops and the band shifts to the blue. Blue shifts induced by the rises of the temperature has been reported for deep, so-called "self activated", DAP luminescent bands of several semiconductors: ZnS [55, 56], GaAs [57] or GaN [58]. The blue shift is generally observed when localized complex centers with strong electron-phonon coupling are involved in the luminescence process. The configurational coordinate (CC) model, which takes into account the interaction of such localized center with matrix, predicts a linear shift of the band position with temperature [56]. The relevant blue shift results from thermal occupation of vibrational levels associated with the excited state and thermal quenching is due to radiationless recombination of $e^- - h^+$ pairs.

The acceptor complex center involved in such emissions is usually formed by an acceptor type vacancy adjacent to a donor impurity atom, [55-58] and it is often called "A-center". The emission results from the electron transition from a relatively distant donor to the donor-vacancy acceptor complex [59]. According to theoretical calculations most stables defects in hBN are supposed to be boron vacancy $V_B$ for n-type conditions and nitrogen vacancy $V_N$ for p-type conditions. To the best of our knowledge there are no calculations concerning vacancy-impurity complexes in hBN. Nevertheless, results obtained in cBN shows that $V_B - C_B$ and $V_B - O_N$ complexes form deep acceptors. Such center in association with the shallow $C_B$ or $O_N$ donor may be responsible for DAP transitions at 3.9 eV observed in hBN polycrystalline samples.

**B – Photo-stimulated luminescence and band gap transitions of hBN**

As we have already mentioned, the PSL excitation onset cannot be a measure of bandgap energy. However in combination with PLE spectra, PSL excitation spectra can provide valuable information concerning exciton and band edge energy positions. Below we will discuss the PSL excitation in the framework of relevant processes depicted in Figure 4 Several interesting details can be remarked from the comparison of PLE and PSL excitation spectra displayed on Figure 6.

Several free exciton transitions in hBN merge in the dominant peak of the PLE spectrum (Figure 3c) in the range of photon energies between 5.8 and 6.0 eV.[34] One remarkable feature in Figure 6 is the non-contribution of these excitons to the PSL excitation. In fact, dissociation of excitons is required for a storage of distant charges, which results in PSL. This dissociation is only possible if the gain in energy due to the charge localization is greater than the exciton binding energy. Using this reasoning, we have previously set up the lower limit to the exciton binding energy: $D_e > 0.4$ eV [17]. Two factors additionally contribute to the inhibition of exciton dissociation: its short lifetime and limited spatial extent. We have not observed free exciton luminescence in our hBN polycrystalline sample: the free excitons are rapidly bound to defects where a part of them recombine radiatively. The lifetime of free excitons in hBN is then small that reduces their dissociation probability. In a similar way, small exciton radius decreases the dissociation probability. The non-contribution of the free exciton of hBN to the PSL processes may be an indication of its small radius and tight binding energy.

The PSL excitation spectrum in Figure 6 follows below 5.7 eV the exponential growth of the PLE spectrum of q-DAP luminescence at 5.3 eV[17]. The q-DAP states apparently contribute to the PSL signal that complements the general schema in Figure 4 by the type IV transition corresponding to the direct q-DAP excitation (Equation 4). This fact seems surprising since no free charge carriers are created in such process required for a distant charge trapping. Indeed it can be understood if we consider the q-DAP transition energy as a function of the pair distance. Closer charge pairs possess a higher transition energy and distant pairs a lower. The charge diffusion within the manifold of q-DAP levels below ionization can therefore take place toward more distant and hence long-lived states, which contribute to PSL. The existence of the exchange between q-DAP traps excited according to Equation 4 was suggested in our previous article[17].

In the same publication we have reported a change in the q-DAP population mechanism at 5.7 eV. At this energy the q-DAP ionization takes place following Equations (2)-(3) and results in the regime crossover from Raman to photoluminescence. This can be ionization of either acceptor or donor states whatever is higher lying. Higher-energy photons between 5.7 and 6.0 eV efficiently produce hBN excitons, however they do not appreciably contribute to PSL. This is seen in Figure 6 in both PLE ($E_{lum}$ = 5.3 eV) and PSL spectra. However, another sharp PSL excitation maximum is observed in the high-energy wing of the exciton absorption band at 6.1 eV. We can relate a high efficiency of the distant charge traps population at this energy to the ionization of acceptor or donor states whatever is lower lying. Resuming, we attribute two spectral features observed at $E_{i1} = 5.7$ eV and $E_{i2} = 6.1$ eV in the PSL excitation spectrum to the direct ionization of donor and acceptor levels (or vice versa) involved in the q-DAP luminescence (processes II and III of Figure 4 and Equations 2-3):

$$\begin{aligned} E_{A^-} &= E_g - E_{i1\,or\,i2} \\ E_{D^+} &= E_g - E_{i2\,or\,i1} \end{aligned} \qquad (7)$$

Keeping in mind this assignment, we can estimate the bandgap energy of hBN. Indeed, the energy of the q-DAP luminescence $E_{qDAP}$ = 5.3 eV is given by the relation

$$E_{qDAP} = E_g - E_{A^-} - E_{D^+} \qquad (8)$$

Combining Equations (7) and (8) we then obtain:

$$E_g = E_{i2} + E_{i1} - E_{qDAP} \approx 6.5\,eV \qquad (9)$$

Alternatively, we can assign the spectral feature observed at $E_{i2} = 6.1$ eV to the dissociation of excitons on the high-lying acceptor or donor state of energy $E_{i1} = 5.7$ eV [17]. In framework of this hypothesis the dissociation of lower-energy excitons is energetically forbidden. Indeed, the energy gain from the charge localization on donor or acceptor is $E_{ex} - E_{i1}$, where $E_{ex} \equiv E_{i2}$ is the exciton energy. For the efficient charges separation, this gain in energy has to be higher than the exciton binding energy $E_g - E_{ex}$. This allows the energy balance equation relative to the process threshold:

$$E_g - E_{ex} = E_{ex} - E_{i1}$$
$$E_g = 2E_{i2} - E_{i1} = 6.5 \, eV \quad (10)$$

Interesting, both alternatives in identification of the spectral feature at $E_{i2} = 6.1$ eV result in the same bandgap energy of $E_g = 6.5$ eV.

Our estimation of the hBN bandgap energy is in good agreement with that obtained from all-electron GW calculations recently performed by Arnaud *et al.*[34]. Moreover, taking the exciton transitions at 5.8 eV, we obtain the exciton binding energy of 0.7 eV. This disagrees with the Wannier-type exciton suggested by Watanabe et al. [1] and supports the Arnaud's prediction of the Frenkel-type exciton with strong binding energy [34].

The shape analysis of PLE and PSL excitation spectra supports the obtained value of hBN bandgap energy. On Figure 7 we have plotted the PLE spectra of the bound excitons luminescence ($E_{lum} = 5.5 \, eV$) together with the PLE spectra of broad DAP luminescence ($E_{lum} = 3.73 \, eV$). The PLE ($E_{lum} = 5.5 \, eV$) peak at 5.8 eV fits well the position of the strongest excitonic transitions predicted theoretically,[34, 60, 61] as represented by vertical bar in Figure 7. As well the plateau of the PLE spectra in the energy range between 6.1 and 6.5 eV is related to low-intensity excitonic transitions converging to the dissociation limit. Interestingly, both PLE spectra coincide in the high-energy spectra range above 6.1 eV and follow the PSL spectrum until 6.5 eV. Above 6.5 eV the PSL grows faster while the intensity of the PLE spectra decreases. The PSL is expected to increase at $h\nu > E_g$, where according to Figure 4, the process I efficiently contributes to the distant traps population. This difference in shapes between the PLE and PSL spectra is therefore explained by interband transitions and supports well our bandgap energy estimation $E_g = 6.5$ eV.

Our final remark concerns the shape of PLE spectrum. From a general point of view, as long as the hBN can be considered as optically thin material, the shape of the PLE spectrum is expected to follow the intrinsic absorption spectrum. At higher absorbance, a saturation or even a decrease of PLE intensity appears as an intensification of non radiative recombination processes. Interestingly, we observe the indication of saturation effect around 6.5 eV in good agreement with our bandgap estimation.

# v. Conclusion.

Photoluminescence of hexagonal boron nitride has been studied by means of the time- and energy-resolved photoluminescence spectroscopy methods. Depending on the excitation energy several luminescence bands have been observed. (i) A strongly structured band in the energy range between 4.1 and 3.3 eV is assigned to phonon replicas of an impurity luminescence. (ii) A very broad band ($\Delta E \sim 1$ eV) centered at 3.9 eV is assigned to DAP transitions involving a strongly localized acceptor complex center. Moreover, the intensity ratio between these two emissions strongly depends on excitation photon energy. (iii) q-DAP are responsible for the 5.3 eV emission and (iv) emission of excitons bound to defects is observed at 5.5 eV.

The excitation of samples by high-energy photons above 5.4 eV enables another phenomenon called photostimulated luminescence, which is due to distantly trapped photoinduced charges. PSL is observed in hBN following interband and acceptor/donor-band transitions. Moreover, we show that in contrast to DAP, PSL can also result from a direct q-DAP excitation. The comparison of photoluminescence excitation and PSL spectra allows bandgap energy estimation of 6.5 eV and supports the hypothesis of Frenkel-like exciton in hBN with large binding energy of 0.7 eV. These conclusions support most recent theoretical calculations.[34, 35]

## Acknowledgments

This work has been supported by the IHP-Contract HPRI-CT-1999-00040 of the European Commission. The authors are grateful to G. Stryganyuk for assistance in conducting experiments at HASYLAB (synchrotron DESY) and to P. Jaffrennou and F. Ducastelle for helpful discussions.

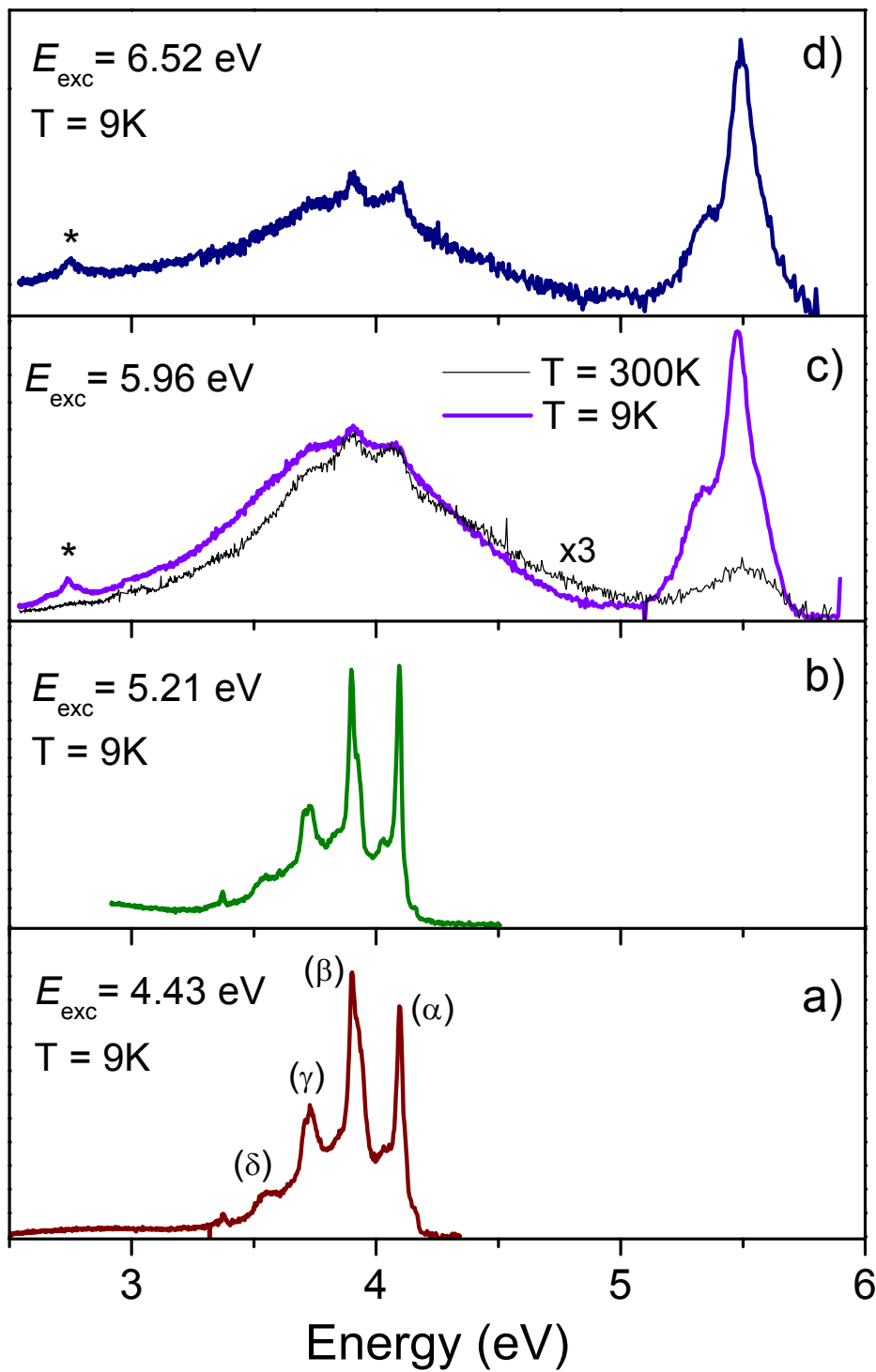

**Figure 1** Luminescence spectra of polycrystalline hBN. The excitation energy is indicated on each spectrum. The phonons replicas are labeled in (a). The star in (c) and (d) indicates the second order of the 5.5 eV band.

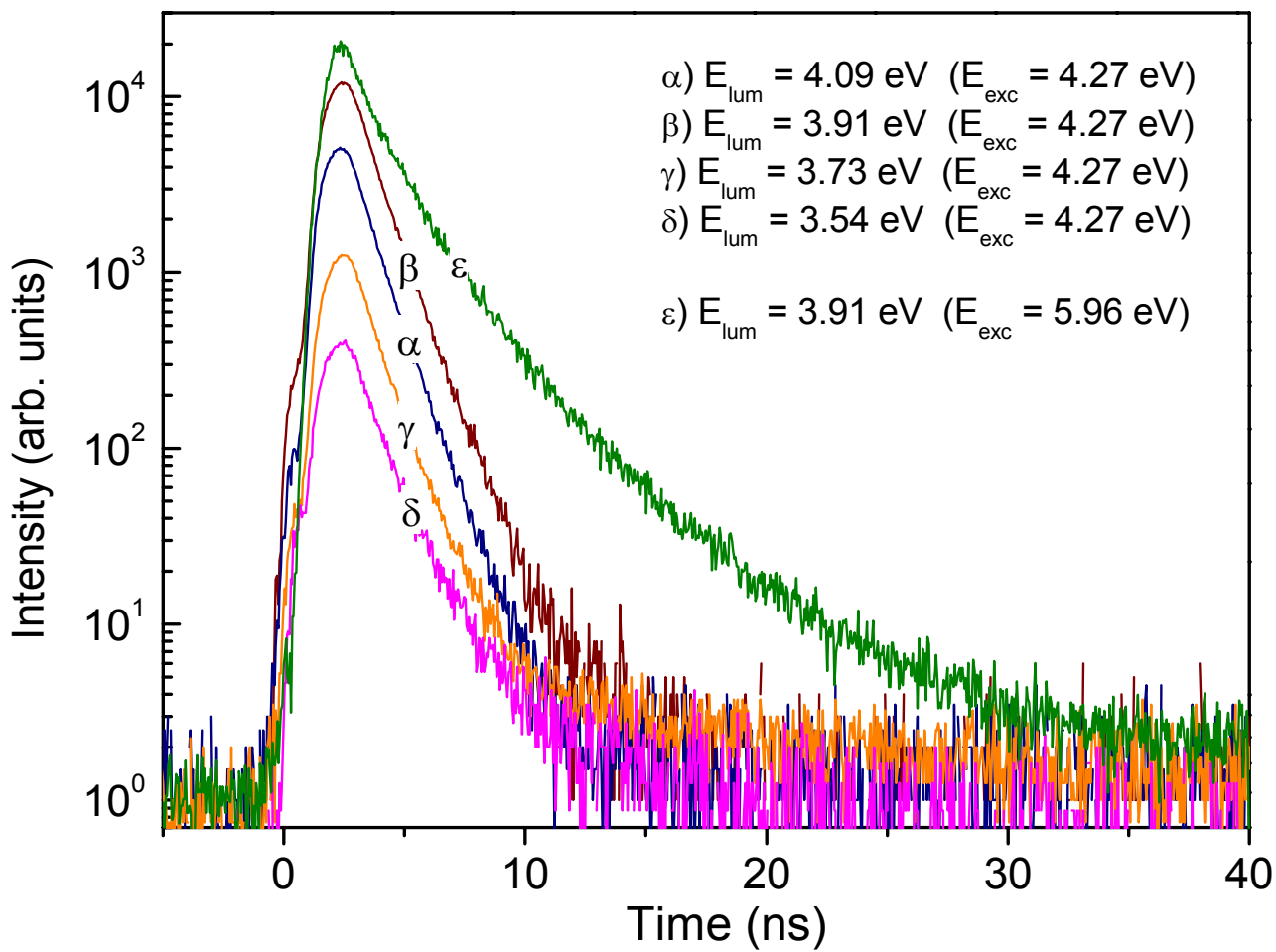

**Figure 2.** Luminescence decay curves at different photon energies. Curves α, β, γ and δ correspond to the peaks of the structured emission. The curve ε corresponds to the broad emission maximum.

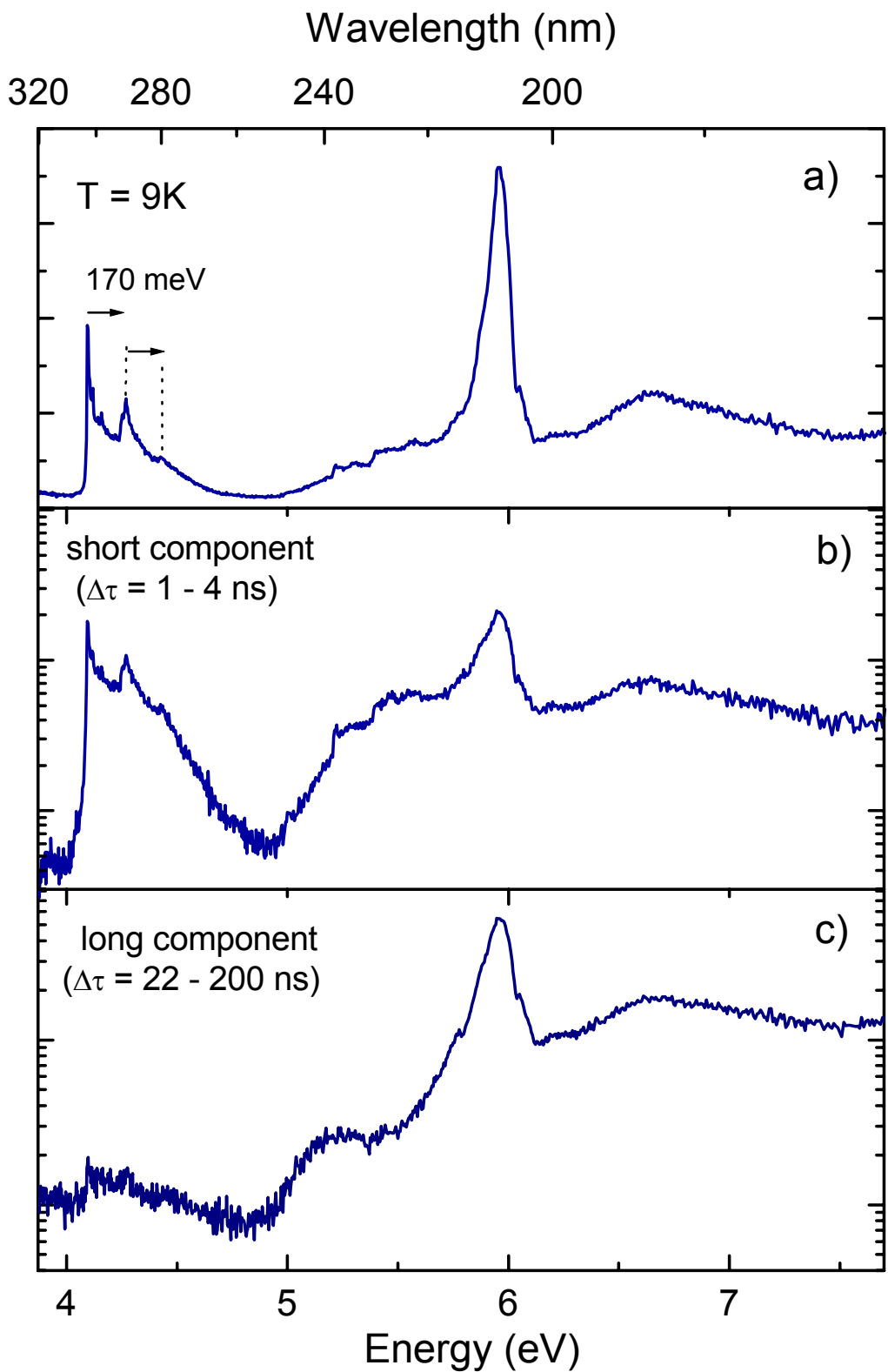

**Figure 3.** PLE spectra ($E_{lum}$ = 3.72 eV) of polycrystalline hBN (a) and its short- (b) and long-lived (c) components.

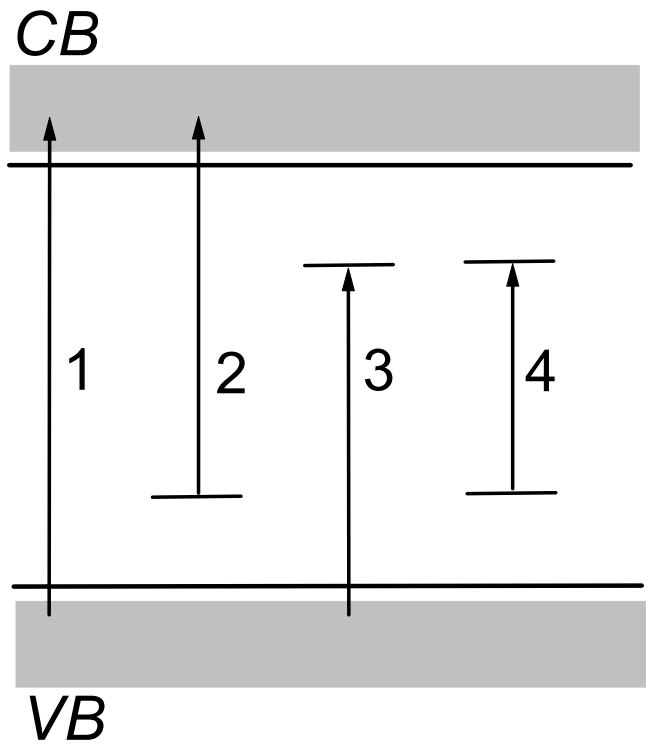

**Figure 4.** Schematic representation of electronic transitions relevant to PSL.

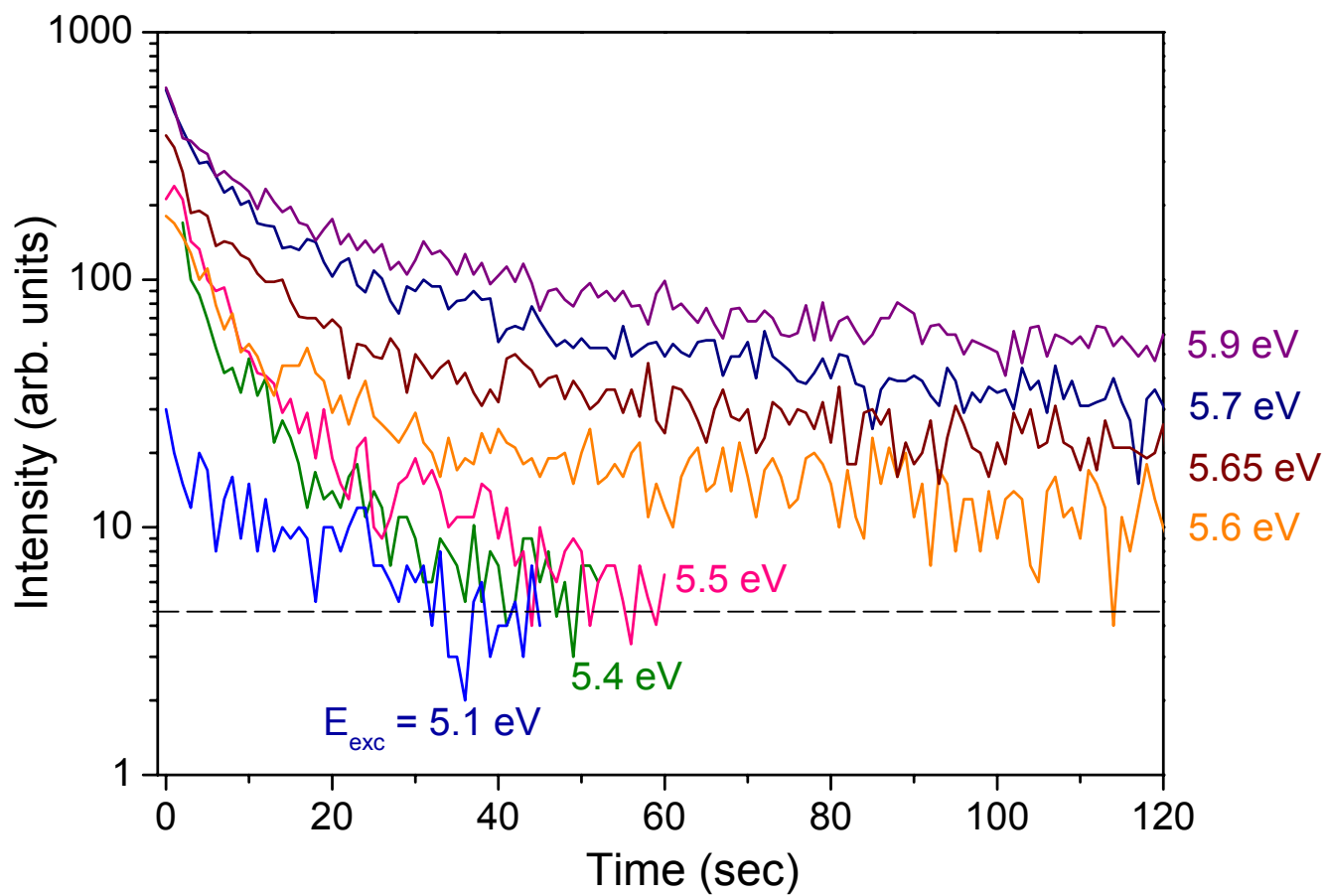

**Figure 5.** Phosphorescence decay curves for different photon excitation energies.

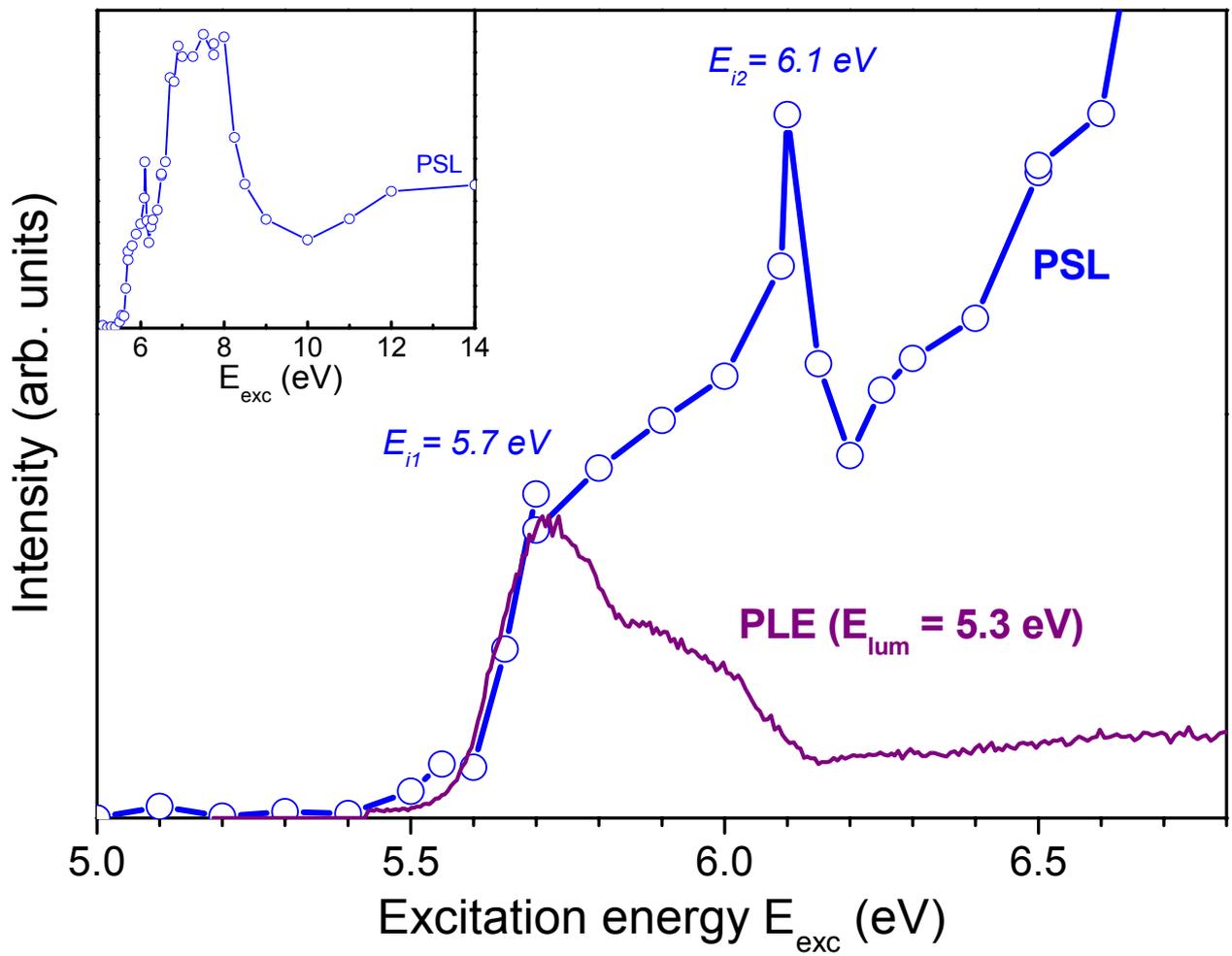

**Figure 6.** PSL excitation spectrum (open circles) and PLE spectrum of q-DAP luminescence (5.3 eV) of polycrystalline hBN. The total measured PSL excitation spectrum is shown in the insert.

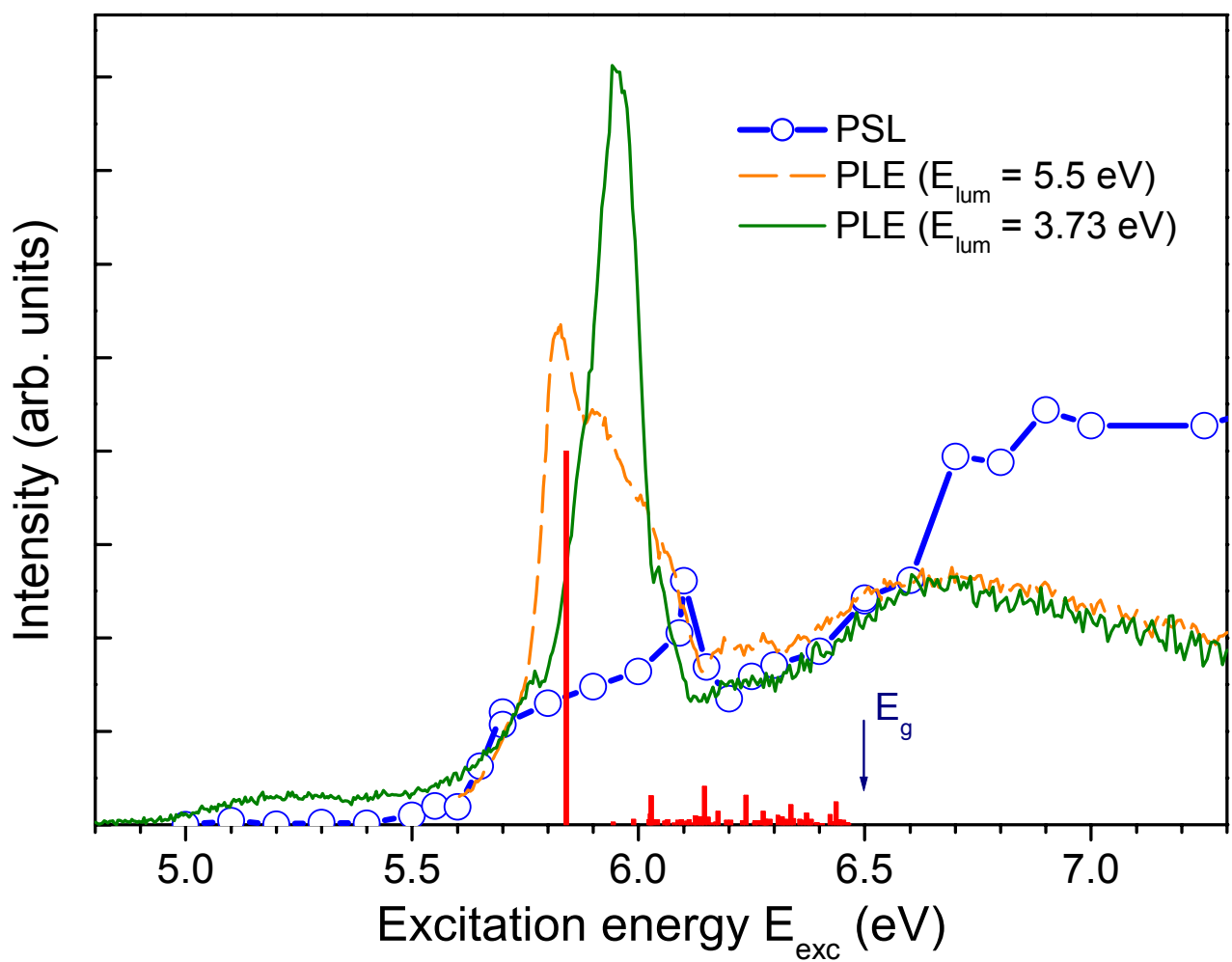

**Figure 7.** PSL excitation spectrum (open circles) and PLE spectra of bound exciton luminescence (5.5 eV, dashed line) and deep DAP luminescence (3.73 eV, solid line). Theoretical oscillator strengths[34, 60, 61] of free exciton transitions in hBN are shown by straight vertical bars.